\documentclass[aps, prb, preprint, showpacs, amssymb, amsmath, superscriptaddress]{revtex4-1}
\usepackage{bm}
\usepackage{times}
\usepackage{graphicx}
\usepackage[colorlinks,citecolor=blue,linkcolor=blue]{hyperref}
\usepackage{array}
\usepackage{float}
\usepackage{colortbl}
\usepackage{multirow}
\usepackage{tabularx}

\begin{document}

\title{Robust Large Gap Quantum Spin Hall Insulators in
 Methyl-functionalized III-Bi Buckled Honeycombs}

\author{Qing Lu}
\affiliation{Institute of Atomic and Molecular Physics, College of Physical Science and Technology, Sichuan University, Chengdu 610065, China}
\affiliation{Beijing National Laboratory for Condensed Matter Physics, Institute of Physics, Chinese Academy of Sciences, Beijing 100190, China}
\author{Busheng Wang}
\affiliation{Beijing National Laboratory for Condensed Matter Physics, Institute of Physics, Chinese Academy of Sciences, Beijing 100190, China}
\author{Xiang-Rong Chen}
\email{xrchen@scu.edu.cn}\affiliation{Institute of Atomic and Molecular Physics, College of Physical Science
and Technology, Sichuan University, Chengdu 610065, China}
\author{Wu-Ming Liu}
\email{wliu@iphy.ac.cn}\affiliation{Beijing National Laboratory for Condensed Matter Physics, Institute of Physics, Chinese Academy of Sciences, Beijing 100190, China}

\begin{abstract}
Based on first-principles calculations, we predict that the methyl-functionalized III-Bi monolayers, namely III-Bi-(CH$_{3}$)$_{2}$ (III=Ga, In, Tl) films, own quantum spin hall (QSH) states with band gap as large as 0.260, 0.304 and 0.843 eV, respectively, making them suitable for room-temperature applications. The topological characteristics are confirmed by $s$-$p_{x,y}$ band inversion, topological invariant $Z_{2}$ number, and the time-reversal symmetry protected helical edge states. Noticeably, for GaBi/InBi-(CH$_{3}$)$_{2}$, the $s$-$p_{x,y}$ band inversion occurred in the progress of spin-orbital coupling (SOC), while for TlBi(CH$_{3}$)$_{2}$, the $s$-$p_{x,y}$ band inversion happened in the progress of chemical bonding. Significantly, the nontrivial topological states in III-Bi-(CH$_{3}$)$_{2}$ films are robust against the mechanical strain and various methyl coverage, making them particularly flexible to substrate choice for device applications. Besides, we find the $h$-BN is an ideal substrate for III-Bi-(CH$_{3}$)$_{2}$ films to realize large gap nontrivial topological states. These findings demonstrate that the methyl-functionalized III-Bi films may be good QSH effect platforms for topological electronic devices design and fabrication in spintronics.
\end{abstract}
\maketitle

\section*{Introduction}
Two-dimensional (2D) topological insulators, also called quantum spin hall (QSH) insulators, have generated great interest in condensed matter physics and materials science due to their novel quantum state and promising applications in spintronics\cite{1,2,3}. The 2D topological insulators are characterized by fully spin-polarized gapless edge states into an insulating bulk, which can achieve non-dissipative electronic transportation. The QSH effect is first proposed by Kane and Mele in graphene\cite{4,5,6}, and experimentally realized in HgTe/CdTe\cite{7,8} and InAs/GaSb\cite{9,10,11} quantum wells. However, due to the very small band gap, the QSH effect only appear at extremely low temperature (-10K). Thus, extensive efforts have been devoted to search new QSH insulators with large band gap. Silicene, germancene\cite{12}, and stanene\cite{13} are verified as QSH insulators sequentially. Then the research emphasis focus to Group-V honeycomb films, in which the Sb (1 1 1)\cite{14} and Bi (1 1 1)\cite{15} monolayers are proved to be QSH insulator intrinsically. Besides, some transition-metal halides\cite{16} and dichalcogenides\cite{17} are also predicted to be large band gap QSH insulators. Recently, the research emphasis further extends to the 2D inversion-asymmetric III-V binary compounds\cite{18,19,20}, in which the appropriate band gap make them suitable for room-temperature applications. Moreover, these III-V films may possess nontrivial topological phenomena ( topological magnetoelectric effects\cite{21}, topological p-n junctions\cite{22}, and surface dependent topological electronic states\cite{23}).

The orbital filtering effect (OFE) plays an important role in tuning the band gap, which has received intense attentions for designing of better QSH insulators. With the effect of halogenation\cite{24}, the band gap of stanene will be enhanced to 0.3 eV. The 2D BiX/SbX (X=H, F, Cl and Br) monolayers are proved to be QSH insulators with extraordinarily large band gaps from 0.32 eV to 1.08 eV\cite{25}. OFE can also be applied to inversion-asymmetric III-V materials\cite{26,27,28,29,30}. By hydrogenation and halogenation, III-Bi films can preferably realize nontrivial topological states with sizable band gaps from 0.25 eV to 0.994 eV\cite{27}. Unfortunately, an experimental work\cite{31} has revealed that hydrogenation and fluorination exhibit quick kinetics, with rapid increase of defects and lattice disorder, which would disrupt their potential applications completely. Recently, the small molecule functionalization is also an effective way to achieve OFE. The ethynyl (C$_{2}$H) functionalized stanene is demonstrated to be a QSH insulator with a band gap as large as 0.3 eV\cite{32}. The methyl (CH$_{3}$) functionalized InBi film (InBiCH$_{3}$)\cite{33} and monolayer Bi (BiCH$_{3}$)\cite{34} are also predicted to be large gap QSH insulators, in which the band gap can be enhanced to 0.29 and 0.934 eV, respectively. GeCH$_{3}$ film has been synthesized experimentally.\cite{35} Unlike the destruction by hydrogenation and fluorination, the methyl functionalization was observed to be much more moderate reaction kinetics, indicating that the methyl is more suitable for surface passivation. Moreover, methyl functionalization can considerably enhance the thermal stability of system at high-temperature\cite{35}. Consequently, it becomes more interesting to realize large band gap QSH insulators in the methyl functionalized III-Bi films.

In this work, we predict a novel family of robust QSH insulators in methyl-functionalized III-Bi buckled honeycombs, namely III-Bi-(CH$_{3}$)$_{2}$ (III=Ga, In, Tl) films, by systematically studying the electronic and topological properties. The III-Bi-(CH$_{3}$)$_{2}$ films are thermal stability, and possess sizable band gap, which may have potential applications at room temperature. Notably, the nontrivial band gap of TlBi(CH$_{3}$)$_{2}$ reaches 0.843 eV. The QSH effect of these films are characterized by $s$-$p_{x,y}$ band inversion, topological invariant $Z_{2}$ = 1, and helical edge states in bulk band gap. In addition, the $s$-$p_{x,y}$ band inversion occurred in the progress of SOC for GaBi/InBi-(CH$_{3}$)$_{2}$, while for TlBi(CH$_{3}$)$_{2}$, the $s$-$p_{x,y}$ band inversion happened in the progress of chemical bonding and the bulk band gap is determined by the splitting strength of SOC effect. Significantly, the QSH states are robust against the mechanical strain and various methyl coverage. Besides, the III-Bi-(CH$_{3}$)$_{2}$ films deposited on $h$-BN substrate is observed to maintain large gap QSH states, which lie within the band gap of $h$-BN substrate.

\section*{Calculation Details}
First-principles calculations are performed using the plane wave basis Vienna ab initio simulation package\cite{36,37}. The electron-ion potential is described by the projector-augmented wave (PAW) method\cite{38}. The electron exchange-correlation functional is approximated by the generalized gradient approximation (GGA) in Perdew-Burke-Ernzerhof (PBE) form\cite{39}. The energy cutoff of the plane wave is set to 500 eV with the energy precision of 10$^{-5}$ eV. The vacuum space is set to at least 20 {\AA} to eliminate the interactions between neighboring slabs. For geometry optimization, the Brillouin zone is sampled by using a $11 \times 11 \times 1$ $\Gamma$-centered Monkhorst-Pack grid\cite{40}, while a $17 \times 17 \times 1$ grid is used for self-consistent calculations. The atomic coordinates were fully optimized until the force on each atom was less than 0.01 eV/{\AA}. The SOC is included in self-consistent electronic structure calculations. The WannierTools package\cite{41} was used to obtain the $Z_{2}$ invariant.It works in the tight-binding framework, which can be generated by another software package Wannier\cite{42}.

\begin{figure*}
\centering
\includegraphics[width= 0.90\textwidth]{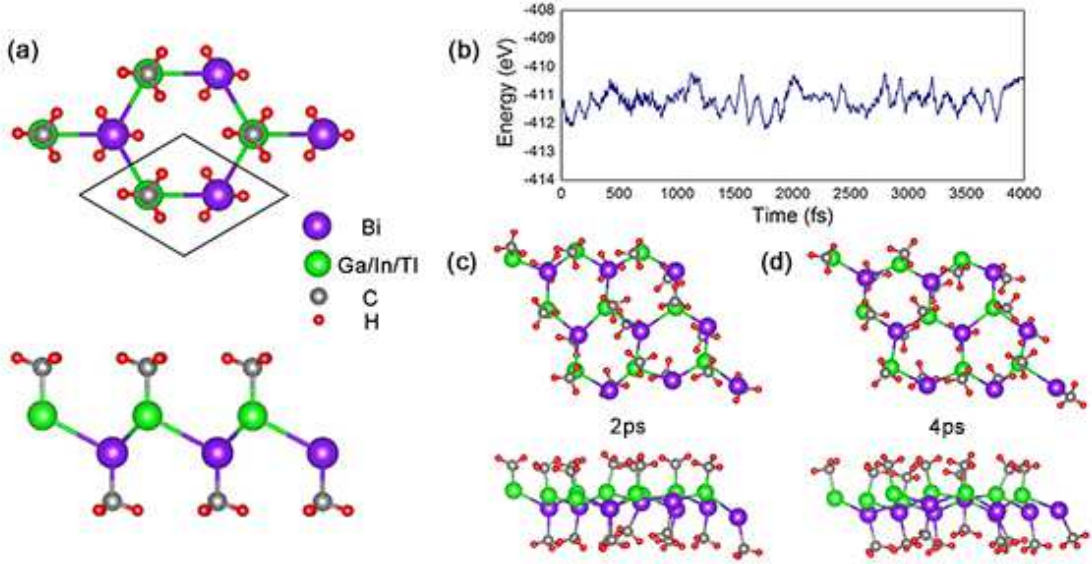}
\caption{ (a) Top and side view of structural representation of III-Bi-(CH$_{3}$)$_{2}$ (III=Ga, In, Tl) films. (b) The total potential energy fluctuation of GaBi(CH$_{3}$)$_{2}$ film during MD simulation at 300 K. (c-d) Snapshots of atomic configurations of GaBi(CH$_{3}$)$_{2}$ film at the end of MD simulation for 2 ps and 4 ps, respectively. The Bi, III, C, and H atoms are highlighted in blue, green, gray, and red spheres, respectively.} \label{fig1}
\end{figure*}

\begin{table*}[!t]\small
\centering \caption{The lattice constant $a$, buckling height $h$, bond length $d_{III-Bi}$, bond length $d_{III-CH_{3}}$, bond length $d_{Bi-CH_{3}}$, global band gap $E_{g}$, band gap at $\Gamma$ point $E_{\Gamma}$ and $Z_{2}$ invariant of III-Bi-(CH$_{3}$)$_{2}$ (III=Ga, In, Tl), respectively.} 
\renewcommand\arraystretch{1.2}  
\begin{tabular}{cccccccccccccccc}
\hline
  Material&$a$({\AA})&$h$({\AA})&$d_{III-Bi}$({\AA})&$d_{III-CH_{3}}$({\AA})&$d_{Bi-CH_{3}}$({\AA})&$E_{g}$(eV)&$E_{\Gamma}$(eV)&$Z_{2}$\\
\hline
GaBi(CH$_{3}$)$_{2}$&4.600&0.856&2.790&2.007&2.286&0.260&0.266&1\\
\hline
InBi(CH$_{3}$)$_{2}$&4.894&0.920&2.972&2.208&2.292&0.304&0.309&1\\
\hline
TlBi(CH$_{3}$)$_{2}$&5.002&0.937&3.037&2.280&2.292&0.843&1.042&1\\
\hline
\end{tabular}
\end{table*}

\section*{Results and Discussion}
\subsection{Crystal structure, stability and electronic properties of the III-Bi-(CH$_{3}$)$_{2}$ films.}
III-Bi-(CH$_{3}$)$_{2}$ (III=Ga, In, Tl) films have a silicene-type hexagonal symmetry structure with one III atom, one Bi atom and two methyl in a unit cell. In comparison with conventional III-Bi monolayers, we saturate the III and Bi atoms with methyl alternating on both sides, as shown in Fig. 1(a). The optimized lattice constants of GaBi(CH$_{3}$)$_{2}$, InBi(CH$_{3}$)$_{2}$, and TlBi(CH$_{3}$)$_{2}$ are 4.600, 4.894, and 5.002 {\AA} with buckling heights of III-Bi layers being 0.856, 0.920, and 0.937 {\AA}, as shown in Table 1. All structural parameters, including lattice constant $a$, buckling height $h$, bond length $d_{III-Bi}$, bond length $d_{III-CH_{3}}$ and bond length $d_{Bi-CH_{3}}$ increase with atomic number Ga, In and Tl. We use following equation to calculate their formation energies, $E_{f}$ = $E$[III-Bi-(CH$_{3}$)$_{2}$] - [$E$(III-Bi) + 2$E$(CH$_{3}$)], where $E$[III-Bi-(CH$_{3}$)$_{2}$] and $E$(III-Bi) are total energies of III-Bi-(CH$_{3}$)$_{2}$ and III-Bi, respectively, while $E$(CH$_{3}$) is chemical potential of methyl. The calculated formation energies for GaBi(CH$_{3}$)$_{2}$,  InBi(CH$_{3}$)$_{2}$ and TlBi(CH$_{3}$)$_{2}$ are -3.75, -3.32 and -2.62 eV, respectively, which are greatly larger than the formation energy of GeCH$_{3}$ (-1.75 eV). Thus, the methyl binds to III-Bi strongly by chemical bonds, displaying highly thermodynamic stability. Considering GeCH$_{3}$ has been synthesized\cite{35}, III-Bi-(CH$_{3}$)$_{2}$ films are expected to be synthesized experimentally.

To examine thermal stability of III-Bi-(CH$_{3}$)$_{2}$ films, we perform ab initio molecular dynamics (MD) simulation with a $3 \times 3 \times 1$ supercell of III-Bi-(CH$_{3}$)$_{2}$ films at 300 K. The fluctuation of total potential energy with time for GaBi(CH$_{3}$)$_{2}$ is illustrated in Fig. 1(b). We found the mean value of total potential energy maintains invariable at whole simulation time. The snapshots of atomic configurations at 2 ps and 4 ps are plotted for GaBi(CH$_{3}$)$_{2}$ film, as shown in Figs. 1(c) and 1(d). There is no structure disruption or structure reconstruction in these systems. We also show the snapshots of atomic configurations of InBi/TlBi(CH$_{3}$)$_{2}$ films at 2 ps in Supplementary Material Fig. S1. Neither structure disruption nor structure reconstruction was occurred in these films. The above results demonstrate that the III-Bi-(CH$_{3}$)$_{2}$ films have good thermal stability and keep its structural integrity at high temperature environment. In addition, we also present the snapshots of atomic configurations of hydrogenation and fluorination III-Bi films at 2 ps, as illustrated in Supplementary Material Fig. S2. Except for GaBiH$_{2}$, InBi/TlBiH$_{2}$ and III-BiF$_{2}$ films appear structure disruption. Thus, compared with hydrogenation and fluorination III-Bi films, methyl-functionalized III-Bi films can considerably enhance the thermal stability at high-temperature, which is similar to the results of methyl-functionalized germanane\cite{35}.

\begin{figure*}
\centering
\includegraphics[width= 0.90\textwidth]{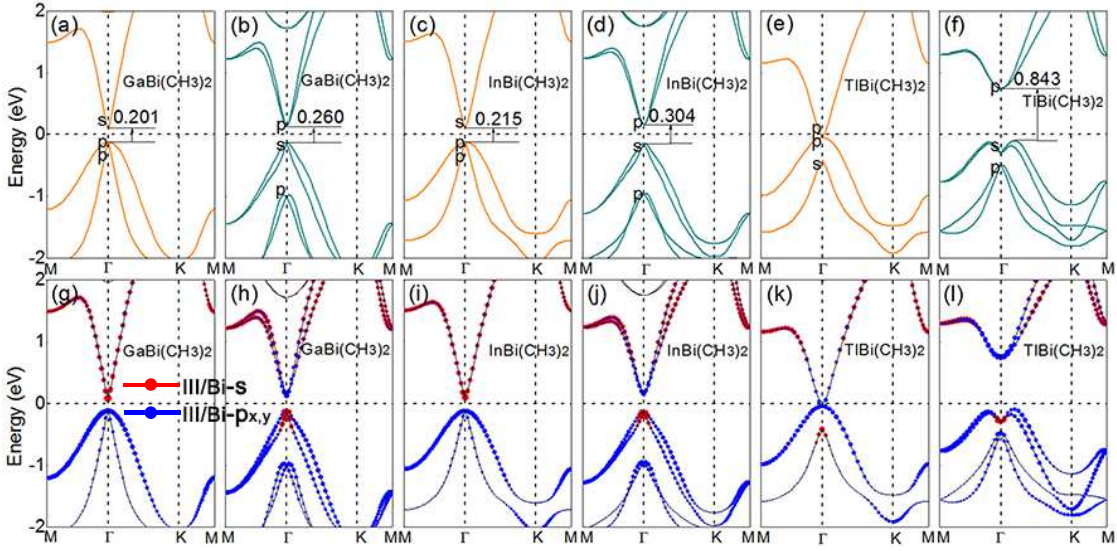}
\caption{The band structures of III-Bi-(CH$_{3}$)$_{2}$ (III=Ga, In, Tl) films: (a) and (b) are the band structures of GaBi(CH$_{3}$)$_{2}$ film. (c) and (d) are same with (a) and (b), but for InBi(CH$_{3}$)$_{2}$ film. (e) and (f) are also same with (a) and (b), but for TlBi(CH$_{3}$)$_{2}$ film. The orange line and dark cyan line represent the band structures without and with SOC, respectively. Insert: the value of the band gap induced by SOC; the main projection of the bands near the Fermi level at $\Gamma$ point. The $s$ and $p$ orbitals indicated in the figure are manily from the III/Bi atoms. The orbitals-resolved band structures of III-Bi-(CH$_{3}$)$_{2}$ (III=Ga, In, Tl) films: (g) and (h) are the orbitals-resolved band structures of GaBi(CH$_{3}$)$_{2}$ film without and with SOC, respectively. (i) and (j) are same with (g) and (h), but for InBi(CH$_{3}$)$_{2}$ film. (k) and (l) are also same with (g) and (h), but for TlBi(CH$_{3}$)$_{2}$ film. The horizontal dashed lines indicate the Fermi level.} \label{fig2}
\end{figure*}

The band structures of III-Bi-(CH$_{3}$)$_{2}$ films are shown in Fig. 2. Without the consideration of SOC, the band structures of GaBi(CH$_{3}$)$_{2}$ and InBi(CH$_{3}$)$_{2}$ present a semiconductor characteristic with a direct band gap situated at $\Gamma$ point, as shown in Figs. 2(a) and 2(c). By presenting the orbitals-resolved band structures in Figs. 2(g) and 2(i), it can be seen that the bands around Fermi level at $\Gamma$ point are mainly derived from one $s$ and two $p_{x,y}$ orbitals of Ga/In and Bi atoms, and the two $p_{x,y}$ orbitals are degenerate at the $\Gamma$ point. For GaBi(CH$_{3}$)$_{2}$ and InBi(CH$_{3}$)$_{2}$, the Fermi level is situated between $s$ and $p_{x,y}$ orbitals, and similar to conventional III-V materials\cite{19,20}, the $s$ orbital is usually situated above $p_{x,y}$ orbitals. Thus, the GaBi(CH$_{3}$)$_{2}$ and InBi(CH$_{3}$)$_{2}$ exhibits a normal band order without considering the SOC. Compared to GaBi(CH$_{3}$)$_{2}$ and InBi(CH$_{3}$)$_{2}$, the band structure of TlBi(CH$_{3}$)$_{2}$ is significantly changed, as shown in Figs. 2(e) and 2(k). It can be found that TlBi(CH$_{3}$)$_{2}$ is gapless semimetal in the absence of SOC, with conduction band minimum and valence band maximum being degenerate at the $\Gamma$ point. Different from the normal band order in GaBi(CH$_{3}$)$_{2}$ and InBi(CH$_{3}$)$_{2}$, the band order of TlBi(CH$_{3}$)$_{2}$ is inverted at the $\Gamma$ point and the two $p_{x,y}$ orbitals is shifted upon $s$ orbital. In former study\cite{27,34}, the $s$-$p_{x,y}$ band inversion usually indicates the formation of QSH phase. It suggests that the TlBi(CH$_{3}$)$_{2}$ could be a QSH insulator once upon turning on a band gap at touch point.

The strong SOC effect has a great influence on band structures of III-Bi-(CH$_{3}$)$_{2}$ films. In GaBi(CH$_{3}$)$_{2}$ and InBi(CH$_{3}$)$_{2}$, the valence band maximum degeneracy is lifted by SOC, as illustrated in Figs. 2(b) and 2(d). Consequently, one of degenerate valence bands is shifted up, while conduction band is shifted down. By presenting orbitals-resolved band structures in Figs. 2(h) and 2(j), it can be found that in GaBi(CH$_{3}$)$_{2}$ and InBi(CH$_{3}$)$_{2}$, the $s$ and $p_{x,y}$ orbitals are inverted with considering the SOC. The $s$-$p_{x,y}$ band inversion indicates the QSH phases in GaBi(CH$_{3}$)$_{2}$ and InBi(CH$_{3}$)$_{2}$. It can be seen in Figs. 2(f) and 2(l), with the effect of SOC, the degenerate two bands is also lifted in TlBi(CH$_{3}$)$_{2}$, with conduction band and valence band shifting upwards and downwards, respectively. It is important to highlight that the effect of SOC in TlBi(CH$_{3}$)$_{2}$ is merely producing a band gap at touch point rather than giving rise to band inversion like GaBi/InBi-(CH$_{3}$)$_{2}$. Actually, a similar case appears in proverbial QSH insulators, such as graphene\cite{4} and silicene\cite{12}, where the SOC effect also does not alter the band order. Due to the inversion-asymmetry in III-Bi-(CH$_{3}$)$_{2}$, the charge will distribute unevenly, which produces inherent surface dipole moment. Combined with 2D configuration, the SOC effect can generate spin splitting of bands away from  $\Gamma$ point, called Rashba effect\cite{27}. It can be seen in Figs. 2(b), 2(d) and 2(f), the valence band maximum situate slightly off $\Gamma$ point, leading to a significant Rashba spin splitting.

\begin{figure*}
\centering
\includegraphics[width= 0.88\textwidth]{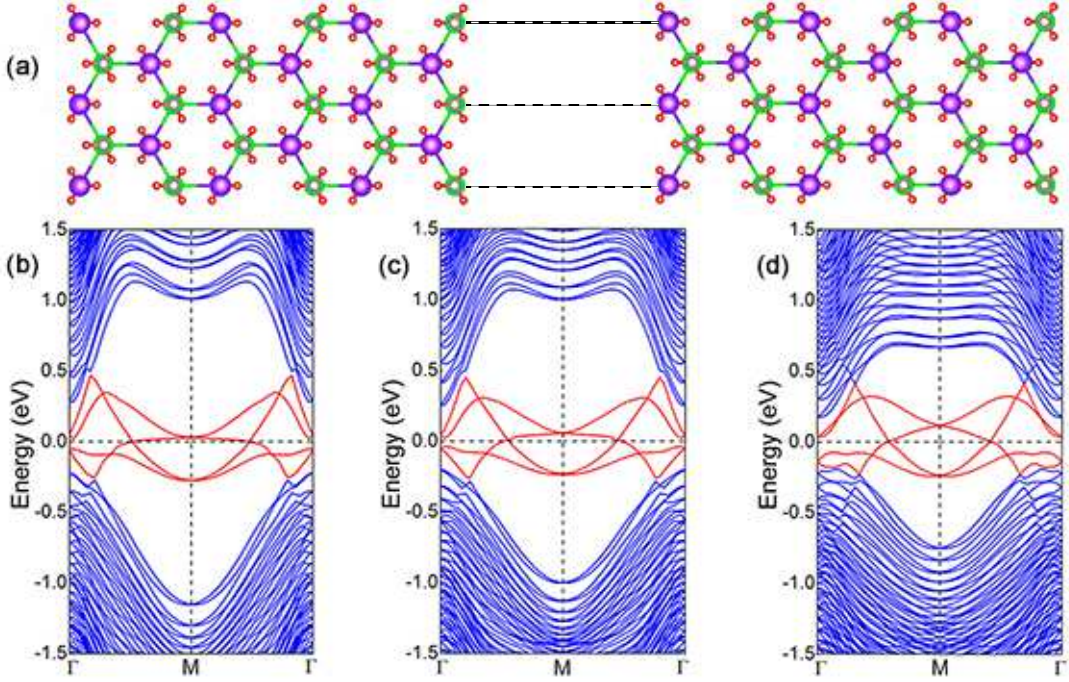}
\caption{Topologically protected edge states: (a) The top view of zigzag nanoribbons for III-Bi-(CH$_{3}$)$_{2}$ films. To avoid the interaction between two sides, the width of these nanoribbons is exceeding 8 nm. (b) The electronic band structure of zigzag nanoribbon for GaBi(CH$_{3}$)$_{2}$ film. (c) and (d) are same with (b), but for InBi(CH$_{3}$)$_{2}$,and  TlBi(CH$_{3}$)$_{2}$ films, respectively. The blue line denotes bulk bands, while the red line represents topologically protected edge bands. The Fermi energy is set to 0 eV.} \label{fig3}
\end{figure*}

\subsection{$Z_{2}$ invariant and topologically protected edge states.}
The topological property can be verified by calculating $Z_{2}$ invariant. Due to inversion-asymmetry in III-Bi-(CH$_{3}$)$_{2}$ films, the method suggested by Fu and Kane\cite{43} cannot be used. Here, we use WannierTools package\cite{41} to calculate the $Z_{2}$ invariant. It can help to classify the topological phase of a given materials by calculating the Wilson loop\cite{44, 45}. In Supplementary Material Fig. S3, we show the wannier charge center evolution along $k_{y}$ for GaBi(CH$_{3}$)$_{2}$, InBi(CH$_{3}$)$_{2}$, and TlBi(CH$_{3}$)$_{2}$ films. The evolution lines cross the arbitrary line parallel to $k_{y}$ odd times, yielding $Z_{2}$ = 1, which indicates that III-Bi-(CH$_{3}$)$_{2}$ films are all QSH insulators.

The existence of topologically protected edge state is the symbol of QSH insulators. To further confirm the nontrivial topological nature of these films, we use the slab model with zigzag nanoribbon to calculate the gapless edge states. For a zigzag nanoribbon, one terminal side is connected by a Bi-CH$_{3}$ chain and other terminal side is connected by a III-CH$_{3}$ chain. The width of these nanoribbons is exceeding 8 nm, which is large enough to ignore the effect between two sides. By the way, the effect of two sides is usually called finite size effect\cite{46}. In addition, to avoid the interaction induced by periodicity, a sufficient vacuum slab is adopted. In Fig. 3, it is clearly shown that for each nanoribbon, there are two sets of edge states corresponding to two opposite sides, which connect conduction and valence bands, and cross linearly at M point. Each nanoribbon displays two Dirac cones, splitting due to asymmetric edges. For a pair of edge states, they cross the Fermi level with odd times from M to $\Gamma$ point. These characteristics strongly indicate that these films are indeed QSH insulators.

\begin{figure*}
\centering
\includegraphics[width= 0.90\textwidth]{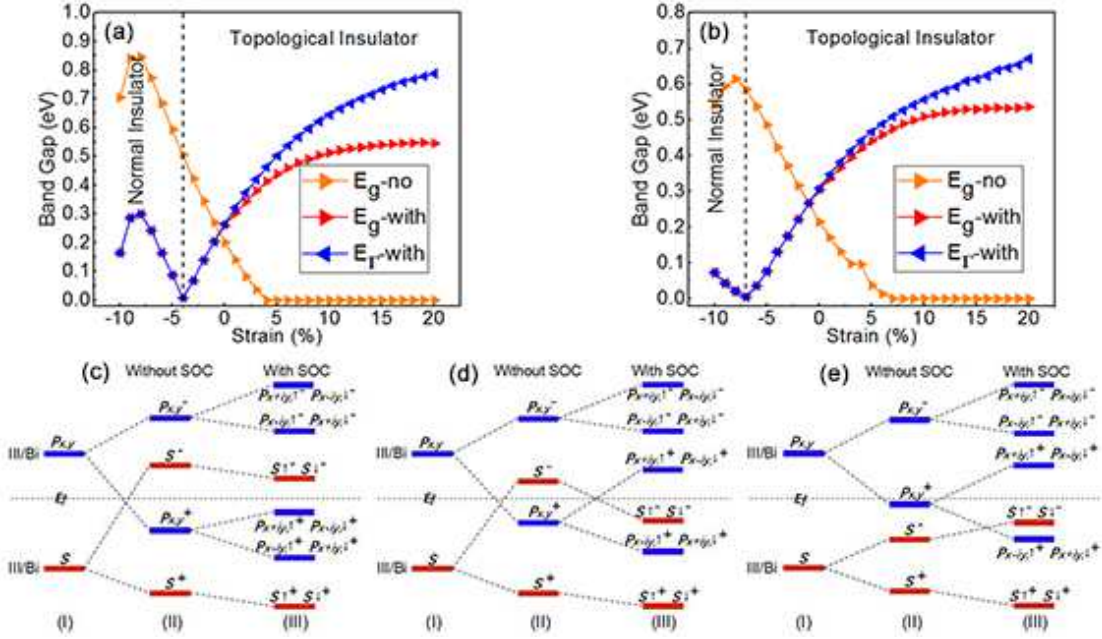}
\caption{The variation of band gap under strain: (a) The variation of band gap E$_{g}$-no, E$_{g}$-with and E$_{\Gamma}$-with as a function of external strain for GaBi(CH$_{3}$)$_{2}$ film. (b)is same with (a), but for InBi(CH$_{3}$)$_{2}$ film. E$_{g}$-no represents the global band gap without SOC, E$_{g}$-with represents the global band gap with SOC and E$_{\Gamma}$-with represents the direct band gap at $\Gamma$ point with SOC. The three types orbital evolution under different strain: (c) No band inversion occurred; (d) band inversion occurred in the progress of spin-orbital coupling SOC; and (e) band inversion happened in the progress of chemical bonding.} \label{fig4}
\end{figure*}

\subsection{The band gap variation and orbital evolution under strain.}
Strain is an efficient way to modulate electronic and topological properties of two-dimensional materials\cite{47,48,49}. The strain in III-Bi-(CH$_{3}$)$_{2}$ films is defined as $\xi = (a - a_{0}) / a_{0}$, where $a (a_{0})$ is lattice constant under strain (equilibrium ) condition. The variation of band gap (E$_{g}$-no, E$_{g}$-with and E$_{\Gamma}$-with) as a function of biaxial strain for GaBi/InBi-(CH$_{3}$)$_{2}$ films is presented in Fig. 4. It can be seen that the nontrivial topological phases exist in GaBi/InBi-(CH$_{3}$)$_{2}$ over a wide strain range, as illustrated in Figs. 4(a) and 4(b). For GaBi(CH$_{3}$)$_{2}$, the E$_{g}$-no decreases monotonically and turns to 0 when strain larger than 4\%. The E$_{g}$-with increases monotonically under tensile strain and tends to be smooth when strain larger than 10\%, and finally reaches a maximum value of 0.546 eV at 20\%. The E$_{\Gamma}$-with also increases monotonically under tensile strain, and reaches a maximum value of 0.789 eV at 20\%. While in compressive strain, the E$_{g}$-with and E$_{\Gamma}$-with are nearly in conformity with each other, and the nontrivial topological state maintains beyond critical point -4\%. If compressive strain increases continuously, a trivial band order appears, forming a normal insulator. For InBi(CH$_{3}$)$_{2}$, the E$_{g}$-no decreases monotonically and turns to 0 when strain larger than 7\%. The E$_{g}$-with and E$_{\Gamma}$-with show a similar tendency, and reaches a maximum value of 0.537 and 0.672 eV at 20\%, respectively. The nontrivial topological states exists beyond critical point -7\%. We also present the variation of band gap (E$_{g}$-with, E$_{\Gamma}$-with) as a function of the biaxial strain for TlBi(CH$_{3}$)$_{2}$ in Supplementary Material Fig. S4(a). It can be seen that the TlBi(CH$_{3}$)$_{2}$ keeps a nontrivial topological state in whole range of strain in our study.

We analyze the orbital evolution of III-Bi-(CH$_{3}$)$_{2}$ films under strain to explain the origin of nontrivial phase. Due to methyl strongly hybridizes with $p_{z}$ orbital at same energy range, it effectually shifts $p_{z}$ orbital away from Fermi level, and thus leaves only $s$ and $p_{x,y}$ orbitals of III and Bi atoms near the Fermi level. In Figs. 4(c) and 4(d), the chemical bonding between III and Bi atoms make the $s$ and $p_{x,y}$ orbitals split into the bonding and antibonding states, which we denote as $|s^{\pm}\rangle$ and $|$$p_{x,y}$$^{\pm}\rangle$, with the subscript $\pm$ representing the bonding and antibonding states. Without considering the SOC effect, the bands around the Fermi level are primarily derived from $|s^{-}\rangle$ and $|$$p_{x,y}$$^{+}\rangle$, and $|s^{-}\rangle$ is located above $|$$p_{x,y}$$^{+}\rangle$, which owns a normal band order. After considering SOC effect, the $|$$p_{x,y}$$^{+}\rangle$ further splits into $|$$p_{x+iy,\uparrow}$$^{+} $$p_{x-iy,\downarrow}$$^{+}\rangle$ and $|$$p_{x-iy,\uparrow}$$^{+} $$p_{x+iy,\downarrow}$$^{+}\rangle$, with $|$$p_{x+iy,\uparrow}$$^{+} $$p_{x-iy,\downarrow}$$^{+}\rangle$ shifting up and $|$$p_{x-iy,\uparrow}$$^{+} $$p_{x+iy,\downarrow}$$^{+}\rangle$ shifting down. At $\xi < -4\%$ for GaBi(CH$_{3}$)$_{2}$ and $\xi < -7\%$ for InBi(CH$_{3}$)$_{2}$, the compressive strain results in a shorter bond length, which enhances the splitting strength of bonding and antibonding states, producing a big energy-difference between $|s^{-}\rangle$ and $|$$p_{x,y}$$^{+}\rangle$. Therefore, the $|$$s_{\uparrow}$$^{-} $$s_{\downarrow}$$^{-}\rangle$ is unable to inverse with $|$$p_{x+iy,\uparrow}$$^{+} $$p_{x-iy,\downarrow}$$^{+}\rangle$, showing a trivial band order, as illustrated in Fig. 4c. While, at $-4\% < \xi < 4\%$ for GaBi(CH$_{3}$)$_{2}$ and $-7\% < \xi < 7\%$ for InBi(CH$_{3}$)$_{2}$, a small energy-difference will be emerged, and thus the SOC effect can promote $|$$p_{x+iy,\uparrow}$$^{+} $$p_{x-iy,\downarrow}$$^{+}\rangle$ higher than $|$$s_{\uparrow}$$^{-} $$s_{\downarrow}$$^{-}\rangle$ Fig. 4d, resulting in a band inversion order, which indicates the existence of nontrivial topological state. Finally, at $\xi > 4\%$ for GaBi(CH$_{3}$)$_{2}$ and $\xi > 7\%$ for InBi(CH$_{3}$)$_{2}$, the tensile strain leads to a longer bond length, which results in a weaker $s$-$p_{x,y}$ hybridization, and accordingly a smaller energy-difference between bonding and antibonding states. Consequently, compared to equilibrium structure of GaBi/InBi-(CH$_{3}$)$_{2}$, the $|s^{-}\rangle$ orbital is shifted down while the $|$$p_{x,y}$$^{+}\rangle$ orbital is shifted up, exhibiting that the $|s^{-}\rangle$ orbital locates under $|$$p_{x,y}$$^{+}\rangle$ orbital, as shown in Fig. 4e. Similar to TlBi(CH$_{3}$)$_{2}$, the $|s^{-}\rangle$ and $|$$p_{x,y}$$^{+}\rangle$ orbitals are inverted in the progress of chemical bonding and the bulk band gap is determined by the splitting of $|$$p_{x,y}$$^{+}\rangle$ under SOC effect. The orbitals-resolved band structures of GaBi(CH$_{3}$)$_{2}$ at 5\% tensile strain and InBi(CH$_{3}$)$_{2}$ at 8\% tensile strain are illustrated in Supplementary Material Fig. S5, which shows a inverted band structure in Fig. 4(e). For TlBi(CH$_{3}$)$_{2}$, the orbital evolution also show a Fig. 4(e) type in  whole range of strain, as shown in Supplementary Material Fig. S4(b).

\begin{figure*}
\centering
\includegraphics[width= 0.80\textwidth]{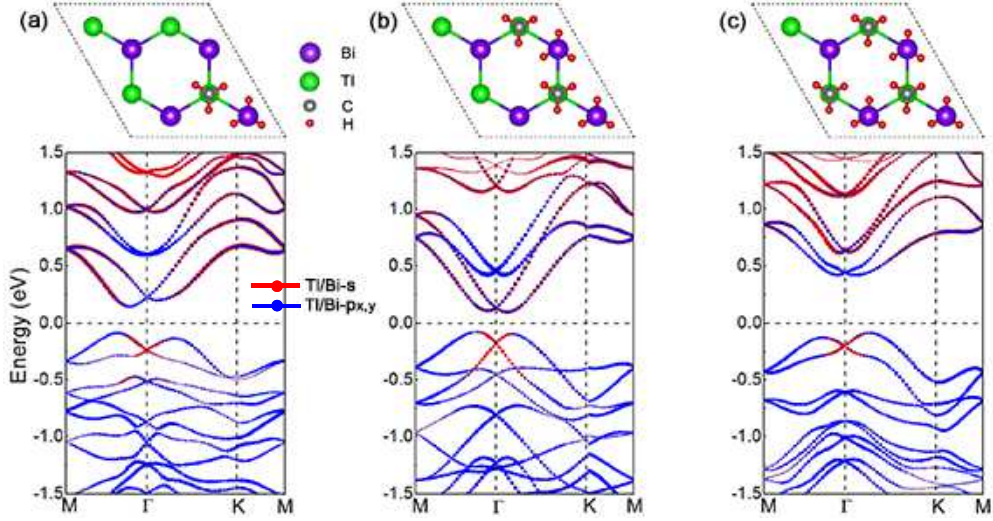}
\caption{The orbitals-resolved band structures of TlBi(CH$_{3}$)$_{2}$ film at various methyl coverage: (a) The structure of TlBi(CH$_{3}$)$_{2}$ film with 25\% methyl coverage and the corresponding orbitals-resolved band structures with SOC. (b) and (c) are same with (a),but for 50\% and 75\% methyl coverage respectively. The red and the blue spheres represent the $s$ and $p_{x,y}$ orbitals, respectively.} \label{fig5}
\end{figure*}

Because the electronic band gap is usually underestimated by the Perdew-Burke-Ernzerhof functional method(PBE)\cite{39}, we adopt the more accurate Heyd-Scuseria-Ernzerhof hybrid functional method (HSE06)\cite{50} to correct it, as shown in Supplementary Material Fig. S6. For GaBi(CH$_{3}$)$_{2}$, the calculated band gap by HSE06 without SOC is 0.694 eV (Supplementary Material Fig. S6(a)), which is larger than the 0.506 eV by PBE in -4\% strain (Fig. 4(a)). Thus, the SOC effect can not induce band inversion and GaBi(CH$_{3}$)$_{2}$ turns to trivial phase. Under 3\% strain, the calculated band gap by HSE06 is 0.437 eV (Supplementary Material Fig. S6(b)), which is smaller than the 0.506 eV by PBE in -4\% strain (Fig. 4(a)). Thus, the SOC effect can induce band inversion and GaBi(CH$_{3}$)$_{2}$ turns to QSH phase again. The InBi(CH$_{3}$)$_{2}$ shows a similar tendency with GaBi(CH$_{3}$)$_{2}$ by HSE06 and turns to QSH phase under 3\% strain (Supplementary Material Figs. S6(c) and S6(d)). While, the TlBi(CH$_{3}$)$_{2}$ shows a gapless semimetal band structure by HSE06 (Supplementary Material Fig. S6(e)) same to PBE and can easily produce a band gap with SOC, which indicates that TlBi(CH$_{3}$)$_{2}$ keeps QSH phase by HSE06. Thus, our PBE-based results in Fig. 4 thus correctly capture evolution of the topological phase with strain, the differences in the critical strain value from normal insulator to topological insulator depending on the particular exchange-correlation functional employed in the computations.

\begin{figure*}
\centering
\includegraphics[width= 0.70\textwidth]{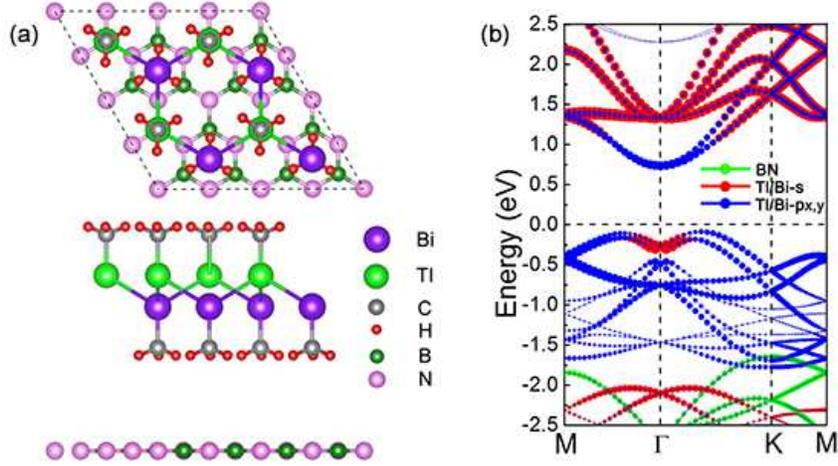}
\caption{(a) The top and side view of TlBi(CH$_{3}$)$_{2}$ deposited on $h$-BN substrate. The Bi, Tl, C, and H atoms are highlighted in blue, green, gray, and red spheres, respectively, while B and N atoms are highlighted in orange and pink shpheres. (b) The orbitals-resolved band structure of TlBi(CH$_{3}$)$_{2}$-$h$-BN heterostructure with SOC. The green sphere represents the contribution of $h$-BN substrate, while the red and the blue spheres represent the $s$ and $p_{x,y}$ orbitals, respectively.} \label{fig6}
\end{figure*}

\subsection{The effects of various methyl coverage and $h$-BN substrate.}
 It can be found that the full methyl functionalized III-Bi films, namely III-Bi-(CH$_{3}$)$_{2}$ (III=Ga, In, Tl), are large gap QSH insulators. To further investigate the robustness of nontrivial topological states, we consider various methyl coverage III-Bi films. $2 \times 2$ supercell of III-Bi films are chosen to simulate 25\%, 50\% and 75\% methyl coverage monolayers, and methyl are placed on alternate sides of III-Bi films, which produces greatly stability. In Figs. 5(a), 5(b) and 5(c), we show the relaxed structures and relevant orbitals-resolved band structures of 25\%, 50\%, and 75\% methyl coverage TlBi films, respectively. Interestingly, all methyl strongly hybridize with $p_{z}$ orbital of Tl and Bi atoms, thus leaves mainly $s$ and $p_{x,y}$ orbitals around the Fermi level. Moreover, the $s$-$p_{x,y}$ band inversion can be observed, which demonstrates that the nontrivial topological states of these films are robustly against chemical bonding effects of environment, making them flexible to substrate option for device applications. We also show the orbitals-resolved band structures for 25\%, 50\% and 75\% methyl coverage GaBi/InBi films, respectively, in Supplementary Material Fig. S7. It can be seen that all structures display a $s$-$p_{x,y}$ band inversion same with methyl coverage TlBi films, indicating the exist of nontrivial topological states.

In terms of device application, it is important to choose a suitable substrate for III-Bi-(CH$_{3}$)$_{2}$ films growth. Since the hydrogenated InBi film on Si(1 1 1) can annihilate its nontrivial topological state\cite{51}, we expect to realize nontrivial topological state by building van der Waals heterostructure for III-Bi-(CH$_{3}$)$_{2}$ films growth. To our knowledge, the $h$-BN is considered as an ideal substrate for its large band gap and high dielectric constant\cite{52}. We establish the TlBi(CH$_{3}$)$_{2}$-$h$-BN heterostructure, as shown in Fig. 6(a), where the lattice mismatch is 2.71\%. After relaxation, the distance between neighboring layers is 2.945 {\AA}, with binding energy of -86 meV, which indicates that the $h$-BN interacts weakly with TlBi(CH$_{3}$)$_{2}$. In Fig. 6(b), we present the orbitals-resolved band structure of TlBi(CH$_{3}$)$_{2}$-$h$-BN heterostructure with SOC. It can be found that the contributions of $h$-BN situate away from the Fermi level and the states near Fermi level are dominantly determined by TlBi(CH$_{3}$)$_{2}$, with an inverted band order. Thus, the TlBi(CH$_{3}$)$_{2}$-$h$-BN heterostructure is a QSH insulator with a large gap of 0.823 eV. Similar to TlBi(CH$_{3}$)$_{2}$-$h$-BN heterostructure, the GaBi/InBi-(CH$_{3}$)$_{2}$-$h$-BN heterostructures are also QSH insulators with band gap as large as 0.499 and 0.377 eV, respectively, as shown in Supplementary Material Figs. S8(b) and S8(c). Other works have investigated the growth of hydrogenated III-V films on Si(111) substrate\cite{51,28}, where one bilayer and two bilayer hydrogenated GaBi film on Si(1 1 1) surface own nontrivial topological states with band gap as large as 0.084 and 0.121 eV, respectively, which are smaller than our obtained band gap. Thus, the $h$-BN is an ideal substrate for III-Bi-(CH$_{3}$)$_{2}$ films growth and devices based on III-Bi-(CH$_{3}$)$_{2}$-$h$-BN heterostructures may be more suitable for room-temperature applications.

\section*{Conclusions}
In summary, we predict a new calss of QSH insulators: III-Bi-(CH$_{3}$)$_{2}$ films with sizable bulk band gap, making them suitable for room-temperature applications. Especially for TlBi(CH$_{3}$)$_{2}$, the nontrivial band gap can reach 0.843 eV, which is larger than many experimentally realized or theoretically predicted topological insulators. The QSH states survive in III-Bi-(CH$_{3}$)$_{2}$ films over a wide range of strain. With the increase of strain, the GaBi/InBi-(CH$_{3}$)$_{2}$ films experience two kinds of band inversion mechanisms. One $s$-$p_{x,y}$ band inversion occurred in the progress of SOC, while another $s$-$p_{x,y}$ band inversion happened in the progress of chemical bonding. Also, the III-Bi-(CH$_{3}$)$_{2}$ films remain QSH states at various methyl coverage, indicating the robustness of its band topology against chemical bonding effects of substrate, which make these films particularly flexible to
substrate choice in device applications. Besides, the III-Bi-(CH$_{3}$)$_{2}$ films on $h$-BN substrate is
observed to maintain large gap QSH states, which demonstrates that the $h$-BN substrate is an ideal substrate for III-Bi-(CH$_{3}$)$_{2}$ films growth. These findings may shed new light on the future design and fabrication of large gap QSH insulators based 2D honeycomb lattices in spintronics.

\section*{Acknowledgments}
This work was supported by the NSAF Joint Fund jointly set up by the National Natural Science Foundation of China and the Chinese Academy of Engineering Physics (Grant No. U1430117) and the Science Challenge Project (Grant No. JCKY2016212A501). W.-M. L. would like to thank the support by the National Key RD Program of China under Grant No. 2016YFA0301500; NSFC under Grants No. 11434015, No. 61227902, No. 11611530676, and No. KZ201610005011; SKLQOQOD under Grant No. KF201403; SPRPCAS under Grants No. XDB01020300, and No. XDB 21030300.

\end{document}